\documentclass[a4paper,11pt]{article}
\usepackage{graphicx}
\usepackage{amsmath}
\usepackage{amssymb}
\newcommand{\ds}{\displaystyle}

\title{Phase Fluctuations and Pseudogap Properties:
Influence of Nonmagnetic Impurities}

\author{
Vadim~M.~Loktev$^1$\footnote{E-mail: {\tt vloktev@bitp.kiev.ua}},
Rachel~M.~Quick$^2$,
Sergei~G.~Sharapov$^{1,2}$\\
$^1${\it Bogolyubov Institute for Theoretical Physics}\\
{of National Academy of Sciences of Ukraine}\\
{\it 14-b Metrologicheskaya Str., Kiev 03143, Ukraine}\\
$^2${\it Department of Physics, University of Pretoria,}\\
{\it Pretoria 0002, South Africa}}

\begin{document}

\maketitle

\begin{abstract}
The presence of nonmagnetic impurities in a 2D ``bad metal'' depresses
the superconducting Berezinskii-Kosterlitz-Thouless transition
temperature, while leaving the pairing energy scale unchanged. Thus
the region of the pseudogap non-superconducting phase, where the
modulus of the order parameter is non-zero but its phase is random,
and which opens at the pairing temperature is substantially bigger
than for the clean system. This supports the premise that fluctuations
in the phase of the order parameter can in principle describe the
pseudogap phenomena in high-$T_c$ materials over a rather wide
range of temperatures and carrier densities. The temperature
dependence of the bare superfluid density is also discussed.
\end{abstract}


\section{Introduction}
The differences between the BCS scenario of superconductivity and
superconductivity in high-$T_c$ materials are well
accepted as experimental facts, although there is no
theoretical consensus about their origin. One of the most
convincing manifestations is the pseudogap, or a
depletion of the single particle spectral weight around the Fermi
level (see for example, \cite{Timusk}).
Another transparent manifestation is the
temperature and carrier density dependencies of the superfluid
density in high-$T_c$ superconductors (HTSC)
\cite{Corson,Panagopoulos,Mesot} which do not fit the
canonical BCS behaviour. In particular the value of the zero
temperature superfluid density is substantially less than
the total density of doped carriers \cite{Emery.conference}.
Currently there are many possible explanations for the unusual
properties of HTSC. One of these is based on the nearly
antiferromagnetic Fermi liquid model \cite{Pines}. Another
explanation, proposed by Anderson, relies on the separation
of spin and charge degrees of freedom. One more approach, which
we will follow in this paper, relates the observed anomalies to
precursor superconducting fluctuations. Different authors argue
that alternative types of superconducting fluctuations are
responsible for the pseudogap, e.g. \cite{Levin}, and the scenario
based on the fluctuations of the order parameter phase suggested
by Emery and Kivelson
\cite{Emery}. The latter scenario we believe to be more relevant
due to low superfluid density and practically 2D character of the
conductivity in HTSC mentioned above.
A microscopic 2D model which elaborates the abovementioned scenario
\cite{Emery} has been studied in the papers
\cite{Gusynin.JETP,Gusynin.new}. The
results obtained show that the condensate phase fluctuations indeed
lead to features which are experimentally observed in HTSC both in
the normal and superconducting states \footnote{Of course,
the contribution from the phase fluctuations
need not be the only or even the major contribution.}.
It is obvious, however, that the present treatment of the phase
fluctuations is incomplete due both to the oversimplified character
of the model and the absence of an explanation for the more recent
advanced experiments \cite{Corson,Panagopoulos,Mesot} on the
temperature and doping dependencies of the superfluid density.
It is well known, however, that the theoretical study of HTSC faces
a lot of computational difficulties due to, for example,
a non-conventional order parameter symmetry, complex
frequency-momentum dependence of the effective quasiparticle
attraction, general form of the quasiparticle dispersion law, etc.
Therefore, in order to obtain analytical results, we have to date only
considered nonretarded $s$-wave pairing in the absence of impurities.
(The attempts to consider the retardation effects were done
in \cite{Turkowski}.)

Nevertheless, a discussion of the effect of impurities seems to be
crucial for a realistic model of the HTSC. Indeed, it is known,
that the itinerant holes in HTSC are created by doping which in turn
introduces a considerable disorder into the system, for instance,
from the random Coulomb fields of chaotically distributed charged
impurities (doped ions) \cite{Pogorelov}. Thus one the purposes of
the present paper is to study the model \cite{Gusynin.JETP,Gusynin.new}
but in the presence of nonmagnetic impurities.

In the theory of ``common metals'' the Fermi energy $\epsilon_{F}$
and the mean transport quasiparticle time $\tau_{tr}$ are
independent quantities which are always assumed to satisfy the
criterion  $\epsilon_{F} \tau_{tr} \gg 1$. In HTSC which are
``bad metals'' \cite{Emery} both $\epsilon_{F}$ and $\tau_{tr}$ are
dependent on the doping and the abovementioned criterion may fail
\cite{Pogorelov}. As an illustration we refer to the remarkable
linear dependence of the normal state resistivity \cite{Timusk}
which implies indeed that $\epsilon_{F} \tau_{tr}$ may be $\sim 1$.
It has been shown \cite{Pogorelov} that for strongly disordered
metallic system superconductivity is absent if the
scattering-to-pairing ratio exceeds a critical value and the
existence of superconductivity in a finite range of doping if this
ratio is not exceeded. We shall not study this case but rather
consider here the more usual (and in some sense simple)
situation, originally studied in the papers
of Anderson \cite{Anderson} and Abrikosov-Gor'kov (AG) \cite{Abrikosov}
(see also \cite{Abrikosov.book}), when the superconducting order is
preexisting and the criterion $\epsilon_{F} \tau_{tr} \gg 1$ is
satisfied.

The Anderson theorem \cite{Anderson} states that in 3D the BCS
critical temperature is unchanged in the presence of nonmagnetic
impurities. However, as discussed in \cite{Gusynin.JETP}, the BCS
critical temperature in 2D is the temperature $T^{\ast}$ at which
the pseudogap opens, while the superconducting transition
temperature transition is the temperature $T_{\rm BKT}$ of the
Berezinskii-Kosterlitz-Thouless (BKT) transition. In contrast to
the former, the latter is defined by a bare superfluid density
(given by the delocalized carriers) which is dependent on
(see below) the concentration of impurities. Thus in 2D case
the superconducting transition temperature, $T_{\rm BKT}$ decreases
with increasing impurity concentration.

Thus, in the model under consideration, the relative size of the
pseudogap phase, $(T^{\ast} - T_{\rm BKT})/T^{\ast}$ is larger in the
presence of impurities than in the clean limit \cite{Gusynin.JETP}.
Therefore it can be observed over a wider range of densities.
The second result obtained is that the value of the
zero temperature superfluid density is less than the total
density of carriers (dopants), so that the presence of impurities may
contribute into this diminishing and in its turn
explain the experimental results \cite{Emery.conference}.
Finally we attempt to interpret qualitatively the recent
experiments on the temperature dependence of the superfluid density
\cite{Corson,Panagopoulos} within our scenario.

A brief overview of the paper follows: In Sec.~\ref{sec:model}
we present the model and derive the main equations.
In Sec.~\ref{sec:comparison} we compare the results obtained
for clean and dirty limits. In particular we compare the values
of $T_{\rm BKT}$, the relative sizes of the pseudogap region and
the values of the bare superfluid density at $T = 0$ and $T$
close to $T_{\rho}$. In Sec.~\ref{sec:superfluid} an attempt
is made to give an explanation for the experimental results
\cite{Corson,Panagopoulos}.

\section{Model and main equations}
\label{sec:model}
Our starting point is a continuum version of the two-dimensional
attractive Hubbard model defined by the Hamiltonian
\cite{Gusynin.JETP,Gusynin.new}:
\begin{eqnarray}
H  = && \int d^2 r \left[
\psi_{\sigma}^{\dagger}(x)
\left(- \frac{\nabla^{2}}{2 m} - \mu \right) \psi_{\sigma}(x)
- V \psi_{\uparrow}^{\dagger}(x) \psi_{\downarrow}^{\dagger}(x)
  \psi_{\downarrow}(x) \psi_{\uparrow}(x) \right.
\nonumber                   \\
&&  \left. + U_{imp} (\mbox{\bf r})
\psi_{\sigma}^{\dagger}(x) \psi_{\sigma}(x) \right],
\label{Hamilton}
\end{eqnarray}
where $x= \mbox{\bf r}, \tau$ denotes the space and imaginary time
variables, $\psi_{\sigma}(x)$ is a fermion field with spin $\sigma
=\uparrow,\downarrow$, $m$ is the effective fermion mass, $\mu$ is
the chemical potential, $V$ is an effective local attraction
constant, and $U_{imp}(\mbox{\bf r})$ is the static potential of
randomly distributed impurities; we take $\hbar = k_{B} = 1$.
The model with the Hamiltonian (\ref{Hamilton}) is equivalent to
a model with an auxiliary BCS-like pairing field which is given in
terms of the Nambu variables as
\begin{eqnarray}
H = && \int d^2 r \left\{
\Psi^{\dagger}(x) \left[\tau_3
\left(- \frac{\nabla^{2}}{2 m} - \mu \right) \right. \right.
\nonumber             \\
&& - \left. \left. \tau_{+}\Phi(x) - \tau_{-}\Phi^\ast(x) +
\tau_{3} U_{imp}(\mbox{\bf r}) \right] \Psi(x)
 +\frac{|\Phi(x)|^2}{V} \right\},
\end{eqnarray}
where $\tau_{\pm} = (\tau_{1} \pm i \tau_{2})/2$, $\tau_{3}$ are
the Pauli matrices, and
$\Phi(x)= V \Psi^\dagger(x)\tau_{-}\Psi(x)=
V \psi_\downarrow(x) \psi_\uparrow (x)$ is the complex order field.
Then the partition function can be presented as a functional
integral over Fermi fields (Nambu spinors) and the auxiliary
fields $\Phi$, $\Phi^{\ast}$.

However, in contrast to the usual method, the modulus-phase
parametrization
$\Phi(x) = \rho(x) \exp[i \theta(x)]$ is necessary for the 2D
model at finite temperatures
(see \cite{Gusynin.JETP,Gusynin.new,Ramakrishnan} and references therein).
To be consistent with this replacement one should also introduce
the spin-charge variables for the Nambu spinors
\begin{equation}
\Psi(x) =\exp[i\tau_3\theta(x)/2]\Upsilon(x)
\label{Nambu.phase}
\end{equation}
with $\Upsilon(x)$ the field operator for neutral fermions.

From the Hamiltonian (\ref{Hamilton}), following \cite{Gusynin.JETP},
one can derive an effective one which is the Hamiltonian of
the classical XY-model
\begin{equation}
H_{XY} = \frac{1}{2} J(\mu, T, \rho) \int d^2 r [\nabla \theta (x)]^2
\label{XY.Hamilton}
\end{equation}
where
\begin{eqnarray}
&& J(\mu, T, \rho) = \frac{T}{16 m \pi^2} \sum_{n = -\infty}^{\infty}
\int d^2 k \mbox{tr} [\tau_3
\langle \mathcal{ G}(i \omega_{n}, \mbox{\bf k}) \rangle]
\nonumber                      \\
&& + \frac{T}{32 m^2 \pi^2} \sum_{n = -\infty}^{\infty}
\int d^2 k k^2 \mbox{tr}
[ \langle \mathcal{ G}(i \omega_{n}, \mbox{\bf k}) \rangle
  \langle \mathcal{ G}(i \omega_{n}, \mbox{\bf k}) \rangle ]
\label{J}
\end{eqnarray}
is the bare (i.e. unrenormalized by the phase fluctuations, but
including pair breaking thermal fluctuations) superfluid stiffness. Here
\begin{equation}
\langle \mathcal{ G}(i \omega_{n}, \mbox{\bf k}) \rangle = -
\frac{ (i \omega_{n} \hat{I} - \tau_{1} \rho ) \eta_{n}
+ \tau_{3} \xi(\mbox{\bf k})}
{(\omega_{n}^{2} + \rho^{2}) \eta_{n}^{2} + \xi^{2}(\mbox{\bf k})}
                       \label{Green.impurity}
\end{equation}
with
\begin{equation}
\eta_{n} = 1 + \frac{1}{2 \tau_{tr} \sqrt{\omega_{n}^{2} + \rho^{2}}},
\qquad
\xi(\mbox{\bf k}) = \frac{\mbox{\bf k}^2}{2m} -\mu,
\qquad \omega_{n} = \pi (2n+1)T
                       \label{eta}
\end{equation}
is the AG \cite{Abrikosov.book} Green's function of neutral fermions
averaged over a random distribution of impurities and written in the
Nambu representation \cite{Evans,Rickayzen.book}.
In writing (\ref{J}) we assumed that
$ \langle \mathcal{ G}(i \omega_{n}, \mbox{\bf k})
\mathcal{ G}(i \omega_{n}, \mbox{\bf k}) \rangle \simeq
\langle \mathcal{ G}(i \omega_{n}, \mbox{\bf k}) \rangle
\langle \mathcal{ G}(i \omega_{n}, \mbox{\bf k}) \rangle$.
This approximation, as shown by AG \cite{Abrikosov.book},
does not change the final result for $J$. Note also that the Green's
function (\ref{Green.impurity}) is valid only when
$\epsilon_{F} \tau_{tr} \gg 1$ which demands the presence of a well
developed Fermi surface, which in turn implies that
$\mu \simeq \epsilon_{F}$. Thus one cannot use the expression
(\ref{Green.impurity}) in the so called Bose limit with
$\mu < 0$ \cite{Gusynin.JETP}. On the other hand, Fermi surface
can be formed even in the bad metals when Ioffe-Regel-Mott criterion
proves to be fulfilled \cite{Pogorelov}.

Substituting (\ref{Green.impurity}) into (\ref{J}), and
using the inequalities $\mu \gg T, \rho$ to extend the limits of
integration to infinity, one arrives at
\begin{equation}
J = \frac{\mu}{4 \pi} + \frac{T \mu }{4 \pi}
\sum_{n=-\infty}^{\infty} \int_{-\infty}^{\infty}
dx \left( \frac{1}{x^2 + (\omega_{n}^{2} + \rho^{2}) \eta_{n}^{2}}
- \frac{2 \omega_{n}^{2} \eta_{n}^{2}}
{[x^2 + (\omega_{n}^{2} + \rho^{2}) \eta_{n}^{2}]^2} \right) .
\label{J.intermediate}
\end{equation}
Eq.(\ref{J.intermediate}) is formally divergent and demands
special care due to the fact that one has to perform the integration over
$x$ before the summation \cite{Abrikosov.book}. Finally one can formally
cancel the divergence \cite{Abrikosov.book} to obtain
\begin{equation}
J = \frac{\mu \rho^{2} T}{4} \sum_{n = -\infty}^{\infty}
\frac{1}{(\omega_{n}^{2} + \rho^{2})
\left[ \sqrt{\omega_{n}^{2} + \rho^{2}} +
\frac{\ds 1}{\ds 2 \tau_{tr}} \right]}.
\label{J.final}
\end{equation}

The temperature of the BKT transition for the XY-model Hamiltonian
(\ref{XY.Hamilton}) is determined by the equation
\begin{equation}
T_{\rm BKT} = \frac{\pi}{2}
J (\mu, T_{\rm BKT}, \rho(\mu, T_{\rm BKT})) .
\label{BKT.temperature}
\end{equation}
The self-consistent calculation of $T_{\rm BKT}$ as a function
of the carrier density $n_f = m \epsilon_{F}/\pi$ requires additional
equations for $\rho$ and $\mu$, which together with
(\ref{BKT.temperature}) form a complete set \cite{Gusynin.JETP}.

When the modulus of the order parameter $\rho(x)$ is treated
in the mean field approximation, the equation for $\rho$
takes the following form \cite{Gusynin.JETP}
\begin{equation}
\frac{2 \rho}{V} = \sum_{n = -\infty}^{\infty}
\int \frac{d^2 k}{(2 \pi)^{2}} \mbox{tr}
[\tau_1 \langle \mathcal{ G}(i \omega_{n}, \mbox{\bf k}) \rangle] ,
\label{rho.gap}
\end{equation}
which formally coincides with the gap equation of the BCS theory.
This coincidence allows one to use the Anderson theorem
\cite{Anderson} which states the dependence of $\rho(T)$ is
the same as that for the clean superconductor and not affected by
the presence of nonmagnetic impurities. It is important to recall
that this theorem is, of course, valid only for the $s$-wave
pairing and small disorder.

There is, however, both physical and mathematical differences
between the gap in the BCS theory and $\rho$
\cite{Gusynin.JETP,Gusynin.new}. In particular the temperature
$T_{\rho}$ which is estimated
by the condition $\rho = 0$  is not related to
the temperature of the superconducting transition, but is
interpreted as the temperature of pseudogap opening $T^{\ast}$
(see details in \cite{Gusynin.JETP}).
The main point, which we would like only to stress here, is that
due to the Anderson theorem \cite{Anderson} the value of
$T_{\rho}$ does not depend on the presence of impurities, while
the temperature $T_{\rm BKT}$, as we will show, is lowered.

The chemical potential $\mu$ is defined by the number equation
\begin{equation}
\sum_{n = -\infty}^{\infty} \int \frac{d^2 k}{(2 \pi)^{2}} \mbox{tr}
[\tau_3 \langle \mathcal{ G}(i \omega_{n}, \mbox{\bf k}) \rangle] = n_f.
\label{number}
\end{equation}
Since we are interested in the high carrier density region
the solution of (\ref{number}) is $\mu \simeq \epsilon_{F}$, so
that in Eqs.(\ref{J.final}) - (\ref{rho.gap}) one can replace
$\mu$ by $\epsilon_{F}$.

Having the temperatures $T_{\rho}$ and $T_{\rm BKT}$ as
functions of the carrier density one can build the phase diagram
of the model \cite{Gusynin.JETP} which consists of three regions.
The first one is the superconducting (here BKT)
phase with $\rho \ne 0$ at $T < T_{\rm BKT}$.
In this region there is algebraic order, or a power law decay of the
$\langle\Phi^{\ast}\Phi\rangle$ correlations.  The second region
corresponds to the pseudogap phase ($T_{\rm BKT} < T < T_{\rho} $).
In this phase $\rho$ is still non-zero but the correlations mentioned
above decay exponentially. The third is the normal (Fermi-liquid)
phase at $T > T_{\rho}$ where $\rho= 0$.  Note that
$\langle \Phi(x) \rangle= 0$
everywhere. While the given phase diagram was derived for the idealized
2D model, there are indications that even for as complicated
layered systems as HTSC the value of the critical temperature for
them may be well estimated using $T_{\rm BKT}$
\cite{Abrikosov.1997,LQSh.1999} even though the
transition undoubtedly belongs to the 3D XY class. It was also pointed
out in \cite{Abrikosov.1997} that a nonzero gap in the one-particle
excitation spectrum can persists even without long-range order.

\section{The comparison of clean and dirty limits}
\label{sec:comparison}

\subsection{Clean limit}
The transport time $\tau_{tr}$ is infinite in the clean limit,
so that
\begin{equation}
J(\epsilon_{F}, T, \rho(\epsilon_{F},T)) =
\frac{\epsilon_{F} \rho^{2} T}{4} \sum_{n = -\infty}^{\infty}
\frac{1}{(\omega_{n}^{2} + \rho^{2})^{3/2}} .
\label{J.clean}
\end{equation}
Near $T_{\rho}$ one can obtain from (\ref{J.clean})
\begin{equation}
J(\epsilon_{F}, T \to T_{\rho}^{-}, \rho \to 0) =
\frac{7 \zeta(3)}{16 \pi^3} \frac{\rho^{2}}{T_{\rho}^{2}}
\epsilon_{F}, \label{J.clean.expand}
\end{equation}
where $\zeta(x)$ is the zeta function. This expression must coincide
with the result from \cite{Gusynin.JETP} which was
derived using the opposite order for the summation and integration.
Inserting the well-known dependence of $\rho(T)$ (see, for example
\cite{Schrieffer})
\begin{equation}
\rho^{2}(T \to T_{\rho}^{-}) = \frac{8 \pi^{2}}{7 \zeta(3)} T_{\rho}^{2}
\left(1 - \frac{T}{T_{\rho}} \right)
\label{rho.dependence}
\end{equation}
and then substituting (\ref{J.clean.expand}) into
(\ref{BKT.temperature}) one obtains the following asymptotic
expression for the BKT temperature in the clean limit for high
carrier densities \cite{Halperin,Gusynin.JETP,Babaev}
\begin{equation}
T_{\rm BKT} = T_{\rho} \left(1 - \frac{4T_{\rho}}{\epsilon_{F}}
\right), \qquad T_{\rm BKT} \lesssim T_{\rho}.
\label{BKT.clean}
\end{equation}

In the high density limit one can also use the equation
\begin{equation}
T_{\rho} = \frac{\gamma}{\pi} \sqrt{2 |\varepsilon_{b}| \epsilon_{F}},
\label{rho.temperature}
\end{equation}
where $\gamma \simeq 1.781$ and $\varepsilon_{b}$ is the energy of
the two-particle bound state in vacuum which is a more convenient
parameter than the four-fermion constant $V$ \cite{Miyake,Gusynin.JETP}.

It is obvious from (\ref{BKT.clean}) and (\ref{rho.temperature})
that the pseudogap region shrinks rapidly for high carrier
densities \footnote{
In 2D for $s$-wave pairing the high density limit
is in fact equivalent the weak coupling BCS limit.}
and one may ask (see, for example,
\cite{Randeria}) whether this scenario can explain the pseudogap
anomalies which are observed over a wide range of temperatures and
carrier densities, since in the clean limit the relative size of
the pseudogap region $(T_{\rho} - T_{\rm BKT})/T_{\rho}$ is, for
instance, less than $1/2$ when the dimensionless ratio
$\epsilon_{F}/|\varepsilon_{b}| \lesssim 128 \gamma^{2}/\pi^2 \simeq 41$.
A crude estimate for the dimensionless ratio for optimally doped
cuprates gives
$\epsilon_{F}/|\varepsilon_{b}| \sim 3 \cdot 10^2$ - $10^3$
\cite{Carter} which indicates that in the clean superconductor
the pseudogap region produced by the phase fluctuations is too small.
Of course, all these estimations are qualitative due to the
simplicity of the model.

The value of the bare superfluid density, $n_{s}(T)$ is
straightforwardly expressed via the bare phase stiffness,
$n_s(T) = 4m J(T)$.  In particular, it follows from
(\ref{J.clean}) that $n_s(T=0) = n_f$. This is not surprising
since $n_s(T=0)$ must be equal to the full density $n_f$ for
any superfluid ground state in a translationary invariant system
\cite{Leggett} and the clean system is translationary invariant.
We note, however, as stated above that in HTSC
$n_{s}(T=0) \ll n_f$ \cite{Emery.conference}.
Substituting (\ref{rho.dependence}) into (\ref{J.clean}) one
obtains for $T$ close to $T_{\rho}$ the bare superfluid
density as $n_{s}(T \to T_{\rho}^{-}) = 2 n_{f} (1 - T/T_{\rho})$.
This behaviour of the bare superfluid density is formally the same
as the behaviour of the full superfluid density in the BCS theory.
Nevertheless it is important to remember that the full superfluid
density in the present model undergoes the Nelson-Kosterlitz
jump at $T_{\rm BKT}$ and is zero for $T > T_{\rm BKT}$.
We note that one can probe experimentally both the bare superfluid
density in high-frequency measurements \cite{Corson} and the full
superfluid density in low-frequency measurements
\cite{Panagopoulos}.

\subsection{Dirty limit}

In the dirty limit the quasiparticle transport time $\tau_{tr}$ is
small ($\tau_{tr} \ll \rho^{-1}(T=0)$) so that one can neglect
the radical in the bracket of (\ref{J.final}) \cite{Abrikosov.book}.
The remaining series is easily summed and one obtains for the
bare superfluid stiffness
\begin{equation}
J(\epsilon_{F}, T, \rho(\epsilon_{F},T), \tau_{tr}) =
\frac{\epsilon_{F} \tau_{tr} \rho}{4} \tanh \frac{\rho}{2T} .
\label{J.dirty}
\end{equation}

As explained above, due to the Anderson theorem, the expressions
(\ref{rho.dependence}) for $\rho$ and (\ref{rho.temperature}) for
$T_{\rho}$ remain unchanged in the presence of
impurities. Again substituting (\ref{rho.dependence}) into
(\ref{J.dirty}) one obtains
\begin{equation}
T_{\rm BKT} = T_{\rho}
\left( 1 - \frac{14 \zeta(3)}{\pi^3}
\frac{1}{\epsilon_{F} \tau_{tr}}  \right), \qquad
T_{\rm BKT} \lesssim T_{\rho}.
\label{BKT.dirty}
\end{equation}
One can see that the size of the pseudogap region is now
controlled by the new phenomenological parameter $\tau_{tr}$ which
is an unknown function of $\epsilon_{F}$ for HTSC.
The experimental data \cite{Timusk} suggest that $\tau_{tr}$
is almost independent on doping level in the underdoped region.

It is difficult to obtain more
than a qualitative estimate using Eq.(\ref{BKT.dirty}) since in
its derivation we have assumed that $\epsilon_{F} \tau_{tr} \gg 1$.
In HTSC however, as discussed above (see also \cite{Pogorelov}),
this assumption is not always
justified. Bearing in mind that the dirty limit implies that
the condition $\tau_{tr}^{-1} \gg \rho(T=0) \sim T_{\rho}$ is
satisfied, one can easily see that the value of $T_{\rm BKT}$ for
this case is less than that given by (\ref{BKT.clean}) for the
clean superconductor.
Since impurities are inevitably present in HTSC, phase fluctuations
can in fact give rise to a pseudogap region that is
of comparable size to that experimentally observed.
We note that our arguments are in fact quite similar to that
given in \cite{Halperin} for the best observing conditions
for the BKT physics in superconducting films. However,
in contrast to this paper, the gap opening below $T_{\rho}$
is particularly emphasized here.

While Eq.(\ref{BKT.dirty}) was derived under assumption
$T_{\rm BKT} \lesssim T_{\rho}$ in general case when $T_{\rm BKT}$
can be substantially less than $T_{\rho}$ the self-consistent
equation (\ref{BKT.temperature}) with
$J(\epsilon_{F}, T_{\rm BKT}, \rho(\epsilon_{F},T_{\rm BKT}))$
given by (\ref{J.dirty}) must be solved.
Recall, however, that to make any quantitative estimates, the more
realistic $d$-wave model has to be considered and the inequality
$\epsilon_{F} \tau_{tr} \gg 1$ should not be assumed
\cite{Pogorelov}.

The value of the zero temperature superfluid density is now
given by $n_{s}(T=0) = \pi n_{f} \tau_{tr} \rho \ll n_{f}$
since $\tau_{tr} \rho \ll 1$. This does not contradict the
results of \cite{Leggett} because the system is not
translationary invariant in the presence of impurities
\cite{Randeria.1998}.
Furthermore as one can see, the low value of the superfluid
density in HTSC \cite{Emery.conference} may be related to
the impurities which are inevitably present in HTSC.
Another reason which leads to lowering of the superfluid density
is the presence of lattice which also destroys a continuous
translational invariance.
We note that as was pointed \cite{Chakraverty} quantum fluctuations
also lead to a decrease in the superfluid density.

\section{The temperature dependence of the bare superfluid density}
\label{sec:superfluid}

In this section we try to correlate the temperature dependence of
the observed in-plane resistivity $\rho_{ab}(T)$ with the recently
measured temperature dependence of the bare superfluid density
\cite{Corson}.

For $T > T_{\rm BKT}$ the expression for the bare superfluid
density in the dirty limit (\ref{J.dirty}) can be rewritten in
terms of the in-plane conductivity,
$\sigma = e^2 n_f \tau_{tr} /m$, where $e$ is the charge of electron:
\begin{equation}
J(\sigma(\epsilon_{F},T), \rho(\epsilon_{F},T)) =
\frac{\pi}{4} \frac{\sigma \rho}{e^{2}} \tanh \frac{\rho}{2T} .
\label{J.conductivity}
\end{equation}

The in-plane resistivity, $\rho_{ab} \sim \sigma^{-1}$ in
cuprates has been extensively studied \cite{Timusk}
and its temperature and concentration dependencies must reflect
the pseudogap properties observed in other experiments. One can
say that $\rho_{ab}(T)$ is linear above $T^\ast \simeq T_\rho$
and roughly linear between $T_{\rm BKT}$ and $T_{\rho}$ but with
a lower slope.  Thus in the interval $T_{\rm BKT} < T < T_{\rho}$
the resistance can be approximately written as
$\rho_{ab}(T) = aT + b$, where $a$ and $b$ are functions of
$\epsilon_{F}$  but not of temperature.

\begin{figure}
\resizebox{0.95\textwidth}{!}{%
  \includegraphics[bb=0 355 275 530]{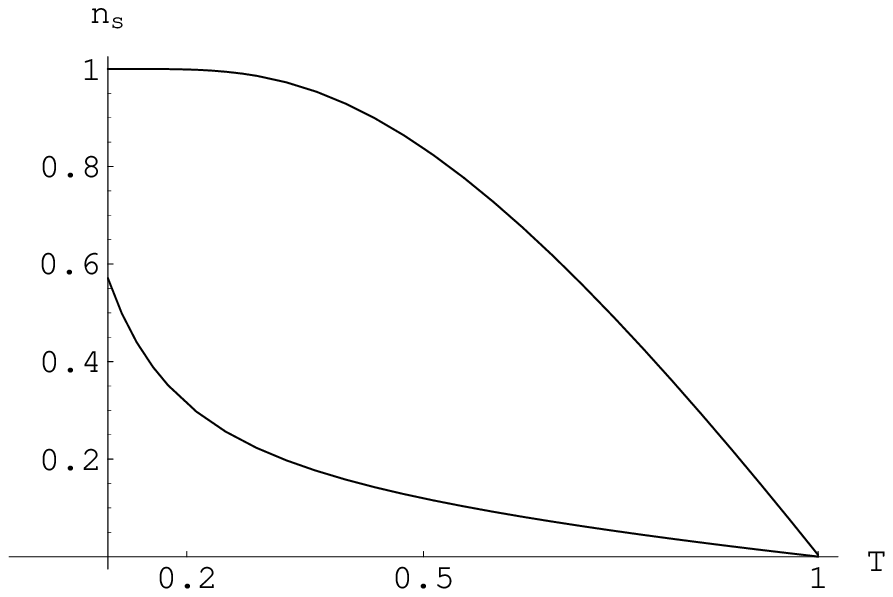}
}
\caption{The behaviour of
$n_s(T)$ in the clean (upper curve) and dirty (lower curve)
limits. The value of $n_s(T)$ is normalized by $n_s(T=0)$
for clean system, $T$ is given in units $T_{\rho}$.}
\label{fig:1}
\end{figure}

Now, substituting $\sigma \sim \rho_{ab}^{-1}(T)$ into
Eq.(\ref{J.conductivity}), one obtains
\begin{equation}
n_{s} (T) \sim \frac{\rho}{aT+b} \tanh \frac{\rho}{2T} .
\label{sup.phenomenological}
\end{equation}
Our estimations based on Eq.(\ref{sup.phenomenological}) are shown
in Fig.~\ref{fig:1}.
One can see that, in contrast to the almost linear BCS
dependence of $n_s(T)$, we have convex behaviour and the superfluid
density becomes zero at $T_{\rho}$. We stress that the curvature of
$n_{s}(T)$ is the result of both the temperature dependence of
$\rho(T)$ and $\sigma(T)$ for $T_{\rm BKT} < T < T_{\rho}$. More
importantly the slope of the curve $n_s(T)$ at $T_{\rho}$ for the
dirty metal is substantially less than for the clean one. The
experiment \cite{Corson} shows the same curvature for $n_{s}(T)$
but indicates that the bare superfluid density disappears at a
lower temperature, $T_{s} < T^{\ast}$.  Since the slope
$dn_{s}(T)/dT$ at $T_{\rho}$ is very small, as predicted
by Eq.(\ref{sup.phenomenological}) and observed experimentally,
the non-zero value of $n_s(T)$ between $T_s$ and $T_\rho$ may
however simply be too small to be experimentally observed. A
definitive answer to this question demands further experiments
and theoretical studies. In particular, $\rho$-fluctuations should
be taken into account \cite{Gusynin.JETP,Gusynin.new}.

One can also comment on the experimentally observed change in
the curvature of the full superfluid density, $\mathcal{ N}_{s}(T)$,
with changing carrier density \cite{Panagopoulos} even though
$\mathcal{ N}_{s}(T)$ cannot be directly linked to the bare superfluid
density $n_{s}(T)$ discussed here.  Although the full superfluid
density disappears above $T_{\rm BKT}$, the curvature present in
the bare superfluid density $n_s(T)$ seems to be retained as a
curvature in the full superfluid density, $\mathcal{ N}_{s}(T)$,
below $T_{\rm BKT}$  \cite{Corson,Panagopoulos}. For low carrier
densities (the underdoped region) the pseudogap region,
$T_{\rm BKT} < T < T_{\rho}$, is larger and therefore the
curvature in $n_s(T)$ is more pronounced. This behaviour seems to
be reflected in the full superfluid density, $\mathcal{ N}_s(T)$
below $T_{\rm BKT}$ \cite{Panagopoulos}. It is important however
to study experimentally and theoretically the concentration
dependence of the bare superfluid density, $n_{s}(T)$ to make a
full comparison with the results from \cite{Panagopoulos}
for $\mathcal{ N}_{s}(T)$.

The experimental data of \cite{Panagopoulos} also show that
$\mathcal{ N}_{s}(T)$ does not display the Nelson-Kosterlitz jump.
This is probably related to the influence of the interlayer
coupling, see Refs. in \cite{Gusynin.new}.

\section{Conclusion}
Since in HTSC the pairing scale $T^\ast$ is different from the
superconducting transition temperature the role of non-magnetic
impurities is not traditional and they in fact define the
superconducting properties of a ``bad metal''. In particular the
presence of nonmagnetic impurities strongly increases the size of
the pseudogap phase originating from the fluctuations of the
phase of the order parameter.  In addition the behaviour of the
superfluid density in the presence of the impurities is closer to
that experimentally observed.

Our results are only qualitative since we have considered the
model with non-retarded $s$-wave attraction and an isotropic
fermion spectrum. However, it is likely that for $d$-wave pairing,
the properties obtained will persist. There is, of course,
a problem why a strong disorder does not destroy $d$-wave
superconductivity when non-magnetic impurities are pair-breaking.
As was suggested by Sadovskii \cite{Sadovskii} even $d$-wave
pairing may persist if coupling is strong enough.
Further studies are necessary, for example, it is important
to explain the concentration dependence
of the superfluid slope, $d n_{s} (T)/dT$ at $T=0$
\cite{Panagopoulos,Mesot}.  Our results also indicate that it would
be interesting to study the BCS-Bose crossover problem
in the presence of impurities, especially in $d$-wave case
\cite{Sadovskii}.

\noindent
We gratefully acknowledge V.P.~Gusynin and Yu.G.~Pogorelov
for many useful discussions.
One of us (S.G.Sh) is grateful to the members of the
Department of Physics of the University of Pretoria for hospitality.
R.M.Q and S.G.Sh acknowledge the financial support of the Foundation
for Research Development, Pretoria.

\newpage

\end{document}